\documentclass[10pt,a4paper]{article}
\usepackage{a4wide}
\usepackage{amssymb}
\usepackage{amsmath,amsthm}
\usepackage{graphicx}
\usepackage{subfigure}
\usepackage{color}
\usepackage[nottoc]{tocbibind}
\usepackage[colorlinks]{hyperref}
\usepackage{url}
\author{
\\
Vincent Leijdekker\\
\footnotesize{ Universiteit van Amsterdam, Korteweg-de Vries Institute for Mathematics}\\
\footnotesize{and}\\
\footnotesize{ABN AMRO, Product Analysis}\\
\\
Peter Spreij\\
\footnotesize{Universiteit van Amsterdam, Korteweg-de Vries Institute for Mathematics }
\\\\
}
\date{}

\numberwithin{equation}{section}

\usepackage{ifthen}
\usepackage{mathrsfs}
\usepackage[footnotesize,bf]{caption}

\renewcommand{\hat}{\widehat}

\newcommand{\A}{\mathscr{A}}
\newcommand{\C}{\mathscr{C}}
\newcommand{\E}{\mathbb{E}}
\newcommand{\F}{\mathcal{F}}
\newcommand{\p}{\mathbb{P}}
\newcommand{\R}{\mathbb{R}}
\newcommand{\dif}{\textrm{d}}

\newcommand{\half}{\frac{1}{2}}

\newcommand{\filtProbSpace}{\left(\Omega,\F,(\F_t)_{t\geq 0},\p\right)}
\newcommand{\probSpace}{\left(\Omega,\F,\p\right)}
\newcommand{\dNLhat}{\left(\dif N_t- \hat \lambda_t\dif t\right)}
\newcommand{\dNLhatO}{\left(\dif N_t- \hat \lambda_t^0\dif t\right)}
\newcommand{\dNLhatOs}{\left(\dif N_s- \hat \lambda_s^0\dif s\right)}

\newcommand{\dNLOs}{\left(\dif N_s- \lambda_s^0\dif s\right)}

\newcommand{\g}{$g(s,t)$}

\newcommand{\dd}[2]
{
    \ifthenelse{#2 < 2} {
        \frac{\partial}{\partial #1}
    }{
        \frac{\partial^#2}{\partial #1^#2}
    }
}

\newcommand{\expec}[2]{\E\left[\left.#1\right\vert #2\right]}

\newtheorem{thm}{Theorem}[section]
\newtheorem{lem}[thm]{Lemma}
\newtheorem{cor}[thm]{Corollary}
\newtheorem{prop}[thm]{Proposition}
\newtheorem{example}[thm]{Example}
\newtheorem{remark}[thm]{Remark}

\hypersetup{pdfauthor = Vincent Leijdekker, pdftitle = Filtering, linkcolor=black, pdfstartview = FitH, citecolor = black,filecolor = black}
\hypersetup{bookmarksnumbered=true, bookmarksopen=true}

\title{Explicit Computations for a Filtering Problem with Point Process Observations with Applications to Credit Risk}
\begin{document}

\maketitle
\begin{abstract}
\noindent
We consider the intensity-based approach for the modeling of default times of one or more companies. In this approach the default times are defined as the jump times of a Cox process, which is a Poisson process conditional on the realization of its intensity. We assume that the intensity follows the Cox-Ingersoll-Ross model. This model allows one to calculate survival probabilities and prices of defaultable bonds explicitly. In this paper we assume that the Brownian motion, that drives the intensity, is not observed. Using filtering theory for point process observations, we are able to derive dynamics for the intensity and its moment generating function, given the observations of the Cox process. A transformation of the dynamics of the conditional moment generating function allows us to solve the filtering problem, between the jumps of the Cox process, as well as at the jumps. Assuming that the initial distribution of the intensity is of the Gamma type, we obtain an explicit solution to the filtering problem for all $t>0$. We conclude the paper with the observation that the resulting conditional moment generating function at time $t$, after $N_t$ jumps, corresponds to a mixture of $N_t+1$ Gamma distributions.\\
\emph{keywords:} Credit Risk, Affine Model, Filtering, Point Process.
\end{abstract}

\newpage
\section{Introduction}
\label{section:intro}
The main goal in credit risk is the modeling of the default time of a company or default times of several companies. The default times are often modeled using the so-called intensity-based approach as opposed to the firm value approach. Here, the default time of a company is modeled as the first jump time of a Cox process, of which the intensity is driven by some stochastic process, e.g.\ Brownian motion, or, in case of more than one company, as consecutive jump times of this Cox process. This approach enables one to calculate survival probabilities, and to price financial derivatives depending on the default of one or more companies, such as defaultable bonds and credit default swaps. The former pays a certain amount at the maturity of the contract in case the underlying company does not default, otherwise it pays a smaller amount, known as the recovery rate. The credit default swap is a form of default insurance, which pays the loss incurred on a default of the underlying company. Overviews of the intensity-based modeling approach can be found in \cite{1998:Lando}, \cite{Giesecke:2004} and \cite{other:ElizaldeI}. In this approach, it is a common assumption that the driving process can be observed, i.e.\ the observed filtration is generated by the Cox process, which can be seen as the default counting process, and by the driving process.\\
In this paper it is assumed that the driving process is \emph{not} observed, and thus only a point process $N_t$ is observed, which introduces a stochastic filtering problem for point processes. In particular the intensity is assumed to follow the Cox-Ingersoll-Ross (CIR) model, where the driving Brownian motion is not observed. General theory for filtering with point processes can be found in Br\'emaud, \cite{book:Bremaud}, for example.
The case in which one assumes a Cox-Ingersoll-Ross model for the intensity and that the initial
value of the intensity is drawn from a Gamma distribution, has also been considered by Frey et al.\ \cite{inproceedings:FreyProsRung}. These authors derive a recursive solution to the filtering problem at jump times of the point process $N_t$.\\
In contrast, in the present paper we also pay attention to the explicit solution to the filtering problem between jump times. We obtain this part of the solution analytically by solving partial differential equations. Furthermore, we consider a different approach to obtain the recursive solution at jump times. By combining these solutions, we obtain a solution for all $t > 0$. It is further observed that the resulting conditional moment generating function at time $t$ corresponds
to a mixture of $N_t+1$ Gamma distributions according to some discrete distribution.\\
The paper is organized as follows: In Section~\ref{section:model} the Cox-Ingersoll-Ross model is discussed and some results for the case of full information are discussed. Next, in Section~\ref{section:filter}, the filtering problem is introduced and some background is given for filtering of point process observations. First, the filtering formulas from \cite{book:Bremaud} are given, and the equations for the conditional intensity and conditional moment generating function are derived. Then, in the second part of Section~\ref{section:filter}, we introduce \emph{filtering by the method of the probability of reference}, and the filtering equations are transformed using the ideas introduced in \cite{article:BoelBenes}. Section~\ref{section:betweenJumps} deals with the filtering problem between the jump times of the point process, given the initial distribution of the intensity at jump times. In Section~\ref{sec:atJumps}, the filtering problem is solved at jump times, and an explicit, recursive solution is obtained, which combines the solutions between and at jumps. Further the resulting conditional moment generating function is analyzed and it is observed that this function agrees with the moment generating function of a mixture of Gamma distributions. The section concludes with an illustration of the mixing probabilities.

\section{Model and Background}
\label{section:model}
One of the main goals in credit risk is the modeling of the default time of a company or the default times of several companies. Over the years two approaches have become popular, the \emph{structural approach} and the \emph{intensity-based approach}. In the structural approach the company value is modeled, for example as a (jump-)diffusion, and the company defaults when its value drops below a certain level. This approach is discussed in more detail in e.g.\ \cite{Giesecke:2004}, \cite{book:BielRut} and \cite{other:ElizaldeII}. In the intensity-based approach the default time is modeled as the first jump of a point process, e.g. a Poisson process or, more general, a Cox process, which is an inhomogeneous Poisson process conditional on the realization of its intensity. In case one considers more than one company, one can model the default times as consecutive jump times of the Cox Process. In \cite{1998:Lando}, \cite{Giesecke:2004} and \cite{other:ElizaldeI} this modeling approach is discussed in more detail, and \cite{article:schonbucher} provides a detailed application. In this paper we focus on the intensity-based approach, where the intensity, $\lambda_t$, of the Cox process has an affine structure, similar to interest term structure models~\cite{article:DuffieKan}. This means that the intensity process $\lambda_t$ follows a stochastic differential equation (SDE) of the form:
\begin{equation}
\label{eq:intensityAffine}
\dif\lambda_t = (a + b\lambda_t)\dif t + \sqrt{c + d\lambda_t}\, \dif W_t,
\end{equation}
for a Brownian Motion $W_t$, with $d>0$. In particular, the focus is on the Cox-Ingersoll-Ross square root (CIR) model, \cite{article:CoxIngersoll}, for the intensity, where the intensity $\lambda_t$ satisfies
\begin{equation}
\label{eq:intensityCIR}
\dif\lambda_t = - \alpha(\lambda_t -\mu_0)\dif t + \beta\sqrt{\lambda_t}\, \dif W_t.
\end{equation}
In \cite[Section 6.2.2.]{book:LambertonLapeyre} one finds parameter restrictions for this model which guarantee positivity of $\lambda_t$. Naturally one should start with a positive initial value $\lambda_0$, and if $\alpha \mu_0 \geq \beta^2/2$, then $\lambda_t$ remains positive with probability one. Note that using the transformation $X_t = \lambda_t + c/d$ and by a reparametrization, $X_t$ satisfies the general SDE (\ref{eq:intensityAffine}), and $\lambda_t$ satisfies (\ref{eq:intensityCIR}). This implies that the general form (\ref{eq:intensityAffine}) and the CIR intensity (\ref{eq:intensityCIR}) are in fact equivalent. Therefore the CIR intensity will be considered in the remainder of this paper.\\
A big advantage of the affine setup is that many relevant quantities in credit risk can be calculated explicitly. Using the formulas from \cite[Section 6.2.2.]{book:LambertonLapeyre} one can, for example, easily calculate the survival probability $\p(\tau > t \vert \F_s)$, with $t > s$ and $\F_t = \F_t^N\vee \F_t^Y$, where the former filtration is generated by the point process $N_t$ and the latter by some process $Y_t$ driving the intensity process.
\begin{example}
\label{ex:survival}
\emph{
Consider, on the filtered probability space $\filtProbSpace$, a random time $\tau > 0$ as the first jump time of a Cox process $N_t$, which intensity follows the CIR model (\ref{eq:intensityCIR}). Further assume that $\F_t = \F_t^N\vee\F_t^W$, where $\F_t^W$ is the filtration generated by the Brownian motion that drives the intensity process. Then one can calculate the survival probability for $t>s$ as
\begin{equation}
\label{eq:survProbFullInfo}
\p(\tau > t \vert \F_s) = 1_{\{\tau > s\}}\expec{e^{-\int_s^t\lambda_udu}}{\F_s^W},
\end{equation}
which follows from formulas in \cite[Chapter 6]{book:BielRut}. Since $\lambda_t$ is a Markov process, one can condition on $\lambda_s$ instead of $\F_s^W$. An application of Proposition 6.2.4.\ from \cite{book:LambertonLapeyre} to (\ref{eq:survProbFullInfo}) yields
\begin{equation}
\label{eq:survivalProb}
\p(\tau > t \vert \F_s) = 1_{\{\tau > s\}}\exp\left(-\alpha\mu_0 \phi(t-s) - \lambda_s \psi(t-s)\right),
\end{equation}
where
\begin{align*}
\phi(t) &= -\frac{2}{\beta^2}\log\left(\frac{2\gamma e^{t(\gamma + \alpha)/2}}{\gamma - \alpha + e^{t\gamma}(\gamma + \alpha)}\right)\\
\psi(t) &= \frac{2\left(e^{\gamma t} - 1\right)}{\gamma - \alpha + e^{t\gamma}(\gamma + \alpha)}\\
\gamma &= \sqrt{\alpha^2 + 2 \beta^2}.
\end{align*}
}\end{example}
\noindent Other relevant quantities, such as the price of a defaultable bond, can also be calculated analytically, under some restrictions on the interest rate, e.g. by posing that the interest rate evolves deterministically. In \cite{inproceedings:FreyProsRung} some of these quantities are considered in more detail.\\
It is a common assumption, which is also followed above, that the filtration $\F_t$ is built up using two filtrations, $\F_t^Y$ and $\F_t^N$, where the first filtration represents the information about the process driving the intensity and the second filtration contains information about past defaults. In this paper it is assumed that the factor $Y$ is not observed which results in a filtering problem of a point process.\\
In the following sections the problem is introduced formally and solved for the case where the intensity follows the CIR model.

\section{The Filtering Problem}
\label{section:filter}
In filtering theory one deals with the problem of partial observations. Suppose that a process $Z_t$ on the probability space $\probSpace$ is adapted to the filtration $\F_t$. Furthermore let the process $Y_t$ be observed, where $Y_t$ is measurable with respect to a smaller filtration $\F_t^Y \subsetneq \F_t$. One is then interested in conditional expectations of the form $\hat Z_t = \E\left[Z_t\vert \F_t^Y\right]$, and one tries to find the dynamics of the process $\hat Z_t$, for instance by showing that it is the solution of a stochastic differential equation.\\
In this section the filtering problem is considered in the case a point process is observed. First some general theory about filtering with point process observations is discussed, and Example~\ref{ex:survival} is continued within the filtering setup. The calculation of the survival probability depends on the conditional moment generating function, for which an SDE is derived. In the second part of this section this equation is transformed in such a way that the filtering problem allows an explicit solution.

\subsection{Filtering Using Point Process Observations}
\label{subsection:filter}
In the case of point process observations the observed process $Y_t$ is equal to the point process $N_t$, with $\F_t$-intensity $\lambda_t$. The process $Z_t$ is assumed to follow the SDE
\begin{equation}
\label{eq:sdeZ}
\dif Z_t = a_t \dif t + \dif M_t,
\end{equation}
for an $\F_t$-progressive measurable $a_t$, with $\int_0^t \vert a_s \vert \dif s < \infty$, and an $\F_t$-local martingale $M_t$. The filtering problem is often cast as the calculation of the conditional expectation $\E[Z_t \vert \F_t^N] =: \hat Z_t$. Using the filtering formulas from \cite[Chapter IV]{book:Bremaud}, a representation of the solution to this filtering problem can be found. In case the (local) martingale $M_t$ and the observed point process have no jumps in common, one has:
\begin{equation}
\label{eq:generalFilter}
\dif \hat Z_t = \hat a_t \dif t + \left( \frac{\hat{Z_{t-}\lambda}_{t-}}{\hat \lambda_{t-}} - \hat Z_{t-} \right)\left(\dif N_t - \hat \lambda_t\dif t\right),
\end{equation}
with $\hat a_t := \E[a_t \vert \F_t^N]$, and $X_{t-} := \lim_{s\uparrow t}X_s$.\\

\begin{example}[Example~\ref{ex:survival} continued]
\label{ex:survival2}
\emph{
When one wants to calculate the survival probability given $\F_t^N$, one has $Z_t = 1_{\{\tau > t\}}$. Combining this with the survival probability in the case of full information, one can calculate the survival probability $\p(\tau > t \vert \F_s^N)$.
\begin{align*}
\p\left(\tau > t \vert \F_s^N\right) &= \expec{\p\left(\tau > t \vert \F_s^N\vee \F_s^W\right)}{\F_s^N}\\
&= 1_{\{\tau > s\}}\exp\left(-\alpha\mu_0 \phi(t-s)\right)\expec{\exp(- \psi(t-s)\lambda_s)}{\F_s^N},
\end{align*}
which can be calculated if an expression for the conditional moment generation function $\hat f(s,t) := \expec{e^{s\lambda_t}}{\F_t^N}$ is available.}
\end{example}
\noindent
The above example illustrates that one can calculate the survival probability if the conditional moment generating function $\hat f(s,t)$ is known. As a first step in the determination of this function, the SDEs of $\hat \lambda_t:=\expec{\lambda_t}{\F_t^N} $ and $\hat f(s,t)$ are determined. First It\^o's formula is used to obtain the SDE for $e^{s\lambda_t}$, where $\lambda_t$ satisfies (\ref{eq:intensityCIR})
\[ \dif e^{s\lambda_t} = \left[\left(-\alpha s+\half s^2 \beta^2\right)\dd{s}{1} e^{s\lambda_t} + s\alpha \mu_0 e^{s\lambda_t}  \right]\dif t + \beta\sqrt{\lambda_t}e^{s\lambda_t}\dif W_t.\]
The filtered versions are obtained by applying formula (\ref{eq:generalFilter}). One obtains for $\hat \lambda_t$
\begin{align}
\label{eq:hatLambdaCIR} \dif \hat \lambda_t &=-\alpha(\hat \lambda_t-\mu_0)\dif t + \left(\frac{\hat{\lambda_t^2}_{-}}{\hat{\lambda}_{t-}}- \hat \lambda_{t-}\right)\dNLhat,
\end{align}
and for $\hat f(s,t)$ one finds
\begin{align}
\label{eq:hatfstCIR} \dif \hat f(s,t) &= \left[\left(-\alpha s+\half s^2 \beta^2\right)\dd{s}{1}\hat f(s,t) + s\alpha \mu_0 \hat f(s,t)\right]\dif t + \left(\frac{\dd{s}{1}\hat f(s,t-) }{\hat \lambda_{t-}}- \hat f(s,t-)\right)\dNLhat.
\end{align}
In general, filtering equations are very difficult, if possible at all, to solve explicitly, since the first equation involves terms with $\hat{\lambda_t^2}$ and the second equation involves combinations of $\hat\lambda_t$ and $\hat f(s,t)$. In order to solve these equations one should also have equations for $\hat {\lambda_t^2}$, but this involves $\hat{\lambda_t^3}$ and so on, assuming that they exist. So instead of trying to solve these equations directly, a different approach is considered in order to find an expression for $\hat f(s,t)$.

\subsection{Filtering by the Method of Probability of Reference}
\label{subsection:filterReference}
In order to solve the problem introduced above, the \emph{filtering by the method of probability of reference} is considered, see \cite[chapter VI]{book:Bremaud} or \cite[Section 2]{article:BoelBenes}. In this approach a second probability measure $\p_0$ and intensity process $\lambda_t^0$ are introduced, such that $N_t - \int_0^t\lambda_s^0ds$ is a martingale with respect to $\F_t$ under $\p_0$. Corresponding to this change of measure one has the likelihood ratio, or density process $\Lambda$, given by
\begin{equation}
\label{eq:likelihoodRatio}
\Lambda_t := \expec{\frac{\dif \p}{\dif \p_0}}{\F_t} = 1 + \int_0^t \Lambda_{s-}\frac{\lambda_{s-} - \lambda_{s-}^0}{\lambda_{s-}^0}\dNLOs.
\end{equation}
This likelihood ratio turns out to be a useful tool to solve the filtering problem for $\hat f(s,t)$. It is known, see e.g.\ \cite{book:Bremaud} for the case $ \lambda_t^0 \equiv 1$, that the filtered version of this likelihood ratio, $\hat \Lambda_t := \expec{\Lambda_t }{\F_t^N}$ follows an equation similar to  (\ref{eq:likelihoodRatio}). One has
\[\hat \Lambda_t = 1 + \int_0^t\hat \Lambda_{s-}\frac{\hat\lambda_{s-} - \hat\lambda_{s-}^0}{\hat\lambda_{s-}^0}\dNLhatOs\]
To solve the filtering problem for $\hat f(s,t)$ an auxiliary function $g(s,t)$ is introduced. It is defined by
\begin{equation}
\label{eq:gst}
g(s,t) := \hat f(s,t) \hat \Lambda_t \exp\left(-\int_0^t\hat\lambda_u^0\dif u\right).
\end{equation}
The exponent is used in order to obtain a simpler SDE of \g. After a solution to this equation has been found, one can obtain $\hat f(s,t)$ by
\begin{equation}
\label{eq:mgfFromGst}
\hat f(s,t) = \frac{g(s,t)}{g(0,t)}.
\end{equation}
It is directly clear that the first and third component of $g(s,t)$ are positive, and from (\ref{eq:likelihoodRatio2}) follows that also the second component is positive, and thus the division in (\ref{eq:mgfFromGst}) is well defined. The solution to the filtering problem is obtained as soon as an expression for $g(s,t)$ is found. In Proposition~\ref{prop:gst} an SDE is derived for $g(s,t)$ for the intensity following the CIR model.
\begin{prop}
\label{prop:gst}
Let $g(s,t)$ be given by (\ref{eq:gst}), then one has, for $t\geq 0$
\begin{equation}
\label{eq:dgstCIR}
\mbox{\emph{d}} g(s,t)= \left[s\mu_0\alpha g(s,t) + \bigg(\frac{1}{2}s^2\beta^2 - s\alpha -1\bigg)\dd{s}{1} g(s,t)\right]\mbox{\emph{d}} t + \left[ \left(\hat\lambda_{t-}^0\right)^{-1}\dd{s}{1} g(s,t-) - g(s,t-)\right]\mbox{\emph{d}} N_t.
\end{equation}
\end{prop}
\begin{proof}
As a first step in proving (\ref{eq:dgstCIR}), one can rewrite the function $g(s,t)$. An alternative expression for $\hat \Lambda_t$ is given by
\begin{equation}
\label{eq:likelihoodRatio2}
\hat \Lambda_t = \prod_{T_n\leq t}\left(\frac{\hat \lambda_{T_n-}}{\hat \lambda_{T_n-}^0}\right)\exp\left(-\int_0^t \left(\hat \lambda_u- \hat \lambda_u^0\right)\dif u\right),
\end{equation}
which can be checked by a direct calculation. From this it is easy to see that
\[g(s,t) \stackrel{(\ref{eq:gst})}{=} \hat f(s,t) \hat \Lambda_t \exp\left(-\int_0^t \hat \lambda_u^0 du\right) = \hat f(s,t) \prod_{T_n\leq t}\left(\frac{\hat \lambda_{T_n-}}{\hat \lambda_{T_n-}^0}\right)\exp\left(-\int_0^t \hat \lambda_u\dif u\right)=: \hat f(s,t) \hat L_t. \]
For $\hat L_t$ one finds the SDE
\[\dif\hat L_t = \frac{\hat L_{t-}\hat \lambda_{t-}}{\hat \lambda_{t-}^0}\dNLhatO - \hat L_{t-}\dif N_t.\]
The SDE in (\ref{eq:dgstCIR}) follows from the product rule
\begin{align*}
\dif g(s,t) &= \hat f(s,t-)\dif \hat L_t + \hat L_{t-}\dif \hat f(s,t) + \Delta \hat f(s,t)\Delta \hat L_t\\
&=\hat f(s,t-)\left(\frac{\hat L_{t-}\hat \lambda_{t-}}{\hat \lambda_{t-}^0}\dNLhatO - \hat L_{t-}\dif N_t\right) + \hat L_{t-}\Bigg(\left[\left(-\alpha s+\half s^2 \beta^2\right)\dd{s}{1}\hat f(s,t) + s\alpha \mu_0 \hat f(s,t)\right]\dif t\\
&\quad + \left(\frac{\dd{s}{1}\hat f(s,t-) }{\hat \lambda{t-}}- \hat f(s,t-)\right)\dNLhat\Bigg) + \left(\frac{\dd{s}{1}\hat f(s,t-) }{\hat \lambda{t-}}- \hat f(s,t-)\right)\left(\frac{\hat L_{t-}\hat \lambda_{t-}}{\hat \lambda_{t-}^0}- \hat L_{t-}\right)\dif N_t.
\end{align*}
Collecting the terms before $\dif t$ and $\dif N_t$, one obtains the equation
\begin{align*}
\dif g(s,t)&=\Bigg(-\hat \lambda_{t} \hat f(s,t) \hat L_t + \left(-\alpha s+\half s^2 \beta^2\right)\dd{s}{1}\hat f(s,t)\hat L_t + s\alpha \mu_0 \hat f(s,t)\hat L_t - \dd{s}{1}\hat f(s,t)\hat L_t + \hat f(s,t)\hat L_t \hat \lambda_t\Bigg)\dif t\\
&\quad+ \Bigg(\frac{\hat f(s,t-)\hat L_{t-}\hat \lambda_{t-}}{\hat \lambda_{t-}^0}-\hat f(s,t-)\hat L_{t-} + \frac{\hat L_{t-}\dd{s}{1}\hat f(s,t-)}{\hat \lambda_{t-}} - \hat L_{t-}\hat f(s,t-) + \frac{\dd{s}{1}\hat f(s,t-) \hat L_{t-}}{\hat \lambda_{t-}^0}\\
&\quad- \frac{\hat f(s,t-)\hat L_{t-}\hat \lambda_{t-}}{\hat \lambda_{t-}^0} - \frac{\dd{s}{1}\hat f(s,t-)\hat L_{t-}}{\hat\lambda_{t-}}+\hat f(s,t-)\hat L_{t-}\Bigg)\dif N_t.
\end{align*}
The result follows by simplifying the last equation.
\end{proof}
\noindent
The right hand side of (\ref{eq:dgstCIR}) depends only on $g(s,t)$ and its partial derivative with respect to $s$. In the next section this equation is solved between jumps, and in section~\ref{sec:atJumps} the equation is solved at jump times of the process $N_t$.

\section{Filtering Between Jumps}
\label{section:betweenJumps}
In the previous sections the filtering problem for point processes has been defined in general terms, and the problem has further been considered for an intensity following the Cox-Ingersoll-Ross model. To solve the filtering problem, one has to solve equation (\ref{eq:dgstCIR}). This equation can be split up into a partial differential equation between jumps of the process $N_t$ and an equation at jumps. In this section the equation between jumps is solved for a general initial condition at time $T>0$. Later on $T$ will be considered as a jump time of $N_t$. Note that an initial condition for $g(s,t)$ is given as
\[g(s,T) = \hat f(s,T) \hat \Lambda_T \exp\left(-\int_0^T\hat\lambda_u^0\dif u\right).\]
For $T = 0$ it follows that
\[
g(s,0) = \hat f(s,0) = \expec{e^{s\lambda_0}}{\F_0^N}\E\left[e^{s\lambda_0}\right],
\]
which is the moment generating function of the intensity at time $t = 0$, since $\F_0^N = \{\emptyset,\Omega\}$.\\
Before the solution to (\ref{eq:dgstCIR}) is found, the specific case is considered, in which all the parameters in the CIR model are set to zero. Albeit a simple example, the analysis of it sheds some light on the approach that will be followed for the general case.
\begin{example}
\label{ex:toy}
\emph{Consider the CIR model in which all the parameters are set to zero. This results in a constant intensity, and thus $d\lambda_t = 0$. The filter equations (\ref{eq:hatLambdaCIR}) and (\ref{eq:hatfstCIR}) reduce to
\begin{align*}
\dif \hat \lambda_t &=  \left( \frac{\hat{\lambda_{t}^2}_-}{\hat{\lambda}_{t-}} - \hat \lambda_{t-}\right)\left(\dif N_t - \hat \lambda_t\dif t\right)\\
 \dif \hat f_t &=  \left(\frac{\hat{\lambda_{t-}f}_{t-}}{\hat{\lambda}_{t-}} - \hat
f_{t-}\right)\left(\dif N_t - \hat \lambda_t\dif t\right).
\end{align*}
The partial differential equation for $g(s,t)$ between jumps reduces to:
\[\dd{t}{1}g(s,t) = - \dd{s}{1} g(s,t).\]
With an initial condition $g(s,T) = w(s)$, one easily finds that the solution to this equation is
\[g(s,t) = w(s - t + T).\]
}
\end{example}
\noindent
In the next section this example is considered once more, where the filter at jump times is considered. We proceed with the case of an intensity following the CIR model.
\begin{prop}
\label{prop:betweenJumps}
Let $\lambda_t$ follow the Cox-Ingersoll-Ross model (\ref{eq:intensityCIR}), and let $g(s,t)$ be given by (\ref{eq:gst}), with an initial condition at time $T$, $g(s,T) = w(s)$. Then, for $T \leq t < T_n$, with $T_n$ the first jump time of $N_t$ after $T$, $g(s,t)$ solves the partial differential equation
\begin{equation}
\label{eq:pdeBetweenJumps}
\dd{t}{1} g(s,t) =  s\mu_0\alpha g(s,t) + \frac{1}{2\rho}(\rho s - \alpha +\tau)(\rho s - \alpha -\tau)\dd{s}{1} g(s,t),
\end{equation}
where  $\rho := \beta^2$ and $\tau := \sqrt{\alpha^2 + 2\beta^2}$. The unique solution to this equation is given by
\begin{align}
\label{eq:solutionBetweenJumps}
\nonumber g(s,t) &= e^{\theta(\alpha -\tau)(t-T)}\left(\frac{2\tau}{\rho s (e^{-\tau (t-T)}-1) +(\tau - \alpha)e^{-\tau(t-T)}+\tau +\alpha}\right)^{2\theta}\\
&\quad \times w\left(\frac{s\left((\alpha+\tau)e^{-\tau(t-T)} + \tau - \alpha\right)+ 2e^{-\tau(t-T)}-2}{\rho s (e^{-\tau(t-T)} -1) + (\tau - \alpha)e^{-\tau(t-T)} +\tau +\alpha }\right),
\end{align}
where $\theta := \frac{\mu_0\alpha}{\rho}$.
\end{prop}
\begin{proof}
The partial differential equation (\ref{eq:solutionBetweenJumps}) for $g(s,t)$ follows directly from Proposition~\ref{prop:gst}, since the jump part of this equation can be discarded.\\
To obtain a solution to this equation a candidate solution is derived by making a number of transformations of the independent variables, until a simple PDE is found, which can be solved explicitly using known techniques. This candidate solution can then be checked to be the solution by calculating its partial derivatives, and inserting these into (\ref{eq:pdeBetweenJumps}).\\
The first transformation is given by
\begin{equation}
\label{eq:transformStoU}
(s,t) \longrightarrow \left(\frac{\rho s - \alpha + \tau}{\rho s - \alpha - \tau},t\right) =: (u,t).
\end{equation}
Instead of $g(s,t)$ one writes $f_1(u,s)$, in terms of the new variable $u$. Using this transformation and the PDE for $g(s,t)$, one can derive a PDE for $f_1(u,t)$, by expressing $s$ in terms of $u$, and expressing the partial derivatives of $g(s,t)$ as partial derivatives of $f_1(u,t)$. The resulting PDE for $f_1(u,t)$ is
\[\dd{t}{1} f_1(u,t) = \mu_0 \alpha \left(\frac{\alpha}{\rho}+\frac{\tau (u+1)}{\rho(u+1)}\right)f_1(u,t) -\tau u \dd{u}{1} f_1(u,t).\]
The second transformation that is used is given by
\[(u,t)\longrightarrow \left(\frac{\log(u)}{\tau},t\right)=:(v,t),\]
where, for the time being, $u$ is tacitly understood to be positive. Instead of the function $f_1(s,t)$, one considers the function $f_2(v,t) := f_1(u,t)$, in terms of the new variable $v$. This transformation results in a partial differential equation for $f_2(v,t)$,
\[\dd{t}{1} f_2(v,t) = \mu_0\alpha \left(\frac{\alpha}{\rho}+\frac{\tau(e^{\tau v}+1)}{\rho(e^{\tau v}-1)}\right)f_2(v,t) - \dd{v}{1} f_2(v,t).\]
The final transformation is given by
\[f_3(v,t) := \log(f_2(v,t)),\]
which results in the PDE for $f_3(v,t)$:
\begin{equation}
\label{eq:pdef3}
\dd{t}{1} f_3(v,t) + \dd{v}{1} f_3(v,t) = \mu_0\alpha \left(\frac{\alpha}{\rho}+\frac{\tau(e^{\tau v}+1)}{\rho(e^{\tau v}-1)}\right).
\end{equation}
This equation can be solved using the method of characteristics, which is explained in chapter 1 and 8 of \cite{book:Chester}, for example. Using this technique the partial differential equation is transformed in an ordinary differential equation by introducing new variables $\xi(v,t)$ and $\zeta(v,t)$. The former is used to replace both $v$ and $t$, and the latter is used to parameterize the initial curve. To be able to solve the PDE an initial condition is required for $f_3(v,t)$. By applying all the previous transformations to the initial condition $g(s,T)=w(s)$, with $t\geq T$, one obtains the initial condition
\[
f_3(v,T) = \log\left(w\left(\frac{e^{\tau v}(\tau + \alpha)+\tau - \alpha}{\rho\left( e^{\tau v} - 1\right)} \right) \right)=: G(v).
\]
Next one has to solve the differential equations
\begin{align*}
\dd{\xi}{1}t(\xi,\zeta)&=1, & \dd{\xi}{1}v(\xi,\zeta) &=1,
\end{align*}
with the initial conditions $t(0,\zeta) = T$ and $v(0,\zeta) = \zeta$. The unique solution to these equations is trivially given by
\begin{align*}
t(\xi,\zeta) &= \xi + T, & v(\xi,\zeta) &= \xi + \zeta.
\end{align*}
Inverting these expressions, yields
\begin{align*}
\xi(v,t) &= t-T, & \zeta(v,t) &= v-t+T.
\end{align*}
Using these transformations, the partial differential equation (\ref{eq:pdef3}) can be transformed into the ordinary differential equation (ODE)
\begin{equation}
\label{eq:df3}
\dd{\xi}{1} f_3(\xi,\zeta) =\mu_0\alpha \left(\frac{\alpha}{\rho}+\frac{\tau(e^{\tau (\xi +\zeta)}+1)}{\rho(e^{\tau (\xi+\zeta)}-1)}\right) = \frac{\mu_0 \alpha(\alpha +\tau)}{\rho}+\frac{2\tau\mu_0\alpha}{\rho(e^{\tau (\xi +\zeta)}-1)}=\theta(\alpha+\tau)  + \frac{2\tau \theta}{e^{\tau(\xi + \zeta)}-1},
\end{equation}
where $\theta = \frac{\mu_0\alpha}{\rho}$. This ordinary differential equation can be solved for the given initial condition $f_3(v,T) = G(v)$. To derive the solution one can start with a candidate solution
\[f_3(\xi,\zeta) = C_1 \log\left(e^{\tau (\xi +\zeta)}-1\right)+C_2\xi + C_3.\]
For $\xi =0$, one has $f_3(0,\zeta) = C_1 \log \left(e^{\tau \zeta}-1\right) + C_3$, and $f_3$ has partial derivative with respect to $\xi$:
\[
\dd{\xi}{1}f_3(\xi,\zeta)= \tau C_1 + \frac{C_1\tau}{e^{\tau (\xi +\zeta)}-1} + C_2.
\]
Using the initial condition $f_3(0,\zeta) = G(\zeta)$, together with the ODE (\ref{eq:df3}), one can find the values of $C_1,\ C_2$ and $C_3$:
\begin{align*}
C_1 &= 2\theta,\\
C_2 &= \theta(\alpha - \tau),\\
C_3 &= G(\zeta) - 2\theta \log \left(e^{\tau \zeta}-1\right).
\end{align*}
This leads to the unique solution
\begin{equation}
\label{eq:solutionf3generalxizeta}
f_3(\xi,\zeta) = \theta(\alpha - \tau)\xi + 2\theta \log\left(e^{\tau(\xi +\zeta)}-1\right) + G(\zeta)-2\theta \log\left(e^{\tau \zeta}-1\right).
\end{equation}
The proof of the uniqueness of this solution is postponed to the end of this proof.\\
Replacing $\xi$ by $t-T$ and $\zeta$ by $v-t+T$, results in
\begin{equation}
\label{eq:solutionf3general}
f_3(v,t) = \theta(\alpha - \tau)(t-T) + 2 \theta \log \left(\frac{e^{\tau v}-1}{e^{\tau (v-t+T)}-1}\right) +\log\left(w\left(\frac{e^{\tau (v-t+T)}(\tau + \alpha)+\tau - \alpha}{\rho\left( e^{\tau (v-t+T)} - 1\right)} \right) \right).
\end{equation}
Next, one obtains a candidate solution for $g(s,t)$, by reversing all the transformations. This gives
\begin{align*}
f_2(s,t) &= e^{\theta(\alpha - \tau)(t-T)}  \left(\frac{e^{\tau v}-1}{e^{\tau (v-t+T)}-1}\right)^{2 \theta} w\left(\frac{e^{\tau (v-t+T)}(\tau + \alpha)+\tau - \alpha}{\rho\left( e^{\tau (v-t+T)} - 1\right)} \right),\\
f_1(s,t) &=e^{\theta(\alpha - \tau)(t-T)}  \left(\frac{u-1}{ue^{-\tau (t-T)}-1}\right)^{2 \theta} w\left(\frac{ue^{-\tau (t-T)}(\tau + \alpha)+\tau - \alpha}{\rho\left( ue^{-\tau (t-T)} - 1\right)} \right).
\end{align*}
By performing the last substitution, (\ref{eq:transformStoU}), an expression for $g(s,t)$ is obtained. One has
\begin{align*}
g(s,t)&=e^{\theta(\alpha - \tau)(t-T)}  \left(\frac{\frac{\rho s - \alpha + \tau}{\rho s - \alpha - \tau}-1}{\frac{\rho s - \alpha + \tau}{\rho s - \alpha - \tau}e^{-\tau (t-T)}-1}\right)^{2 \theta} w\left(\frac{\frac{\rho s - \alpha + \tau}{\rho s - \alpha - \tau}e^{-\tau (t-T)}(\tau + \alpha)+\tau - \alpha}{\rho\left( \frac{\rho s - \alpha + \tau}{\rho s - \alpha - \tau}e^{-\tau (t-T)} - 1\right)} \right)\\
&= e^{\theta(\alpha -\tau)(t-T)}\left(\frac{2\tau}{\rho s (e^{-\tau (t-T)}-1) +(\tau - \alpha)e^{-\tau(t-T)}+\tau +\alpha}\right)^{2\theta}\\
&\quad \times w\left(\frac{s\left((\alpha+\tau)e^{-\tau(t-T)} + \tau - \alpha\right)+ 2e^{-\tau(t-T)}-2}{\rho s (e^{-\tau(t-T)} -1) + (\tau - \alpha)e^{-\tau(t-T)} +\tau +\alpha }\right),
\end{align*}
where it was used that $(\alpha+\tau)(\tau-\alpha)= 2\rho$. By inserting this candidate into equation (\ref{eq:pdeBetweenJumps}), one can check that it indeed is the solution.\\
The last thing to proof is the uniqueness of the solution to equation (\ref{eq:pdeBetweenJumps}). As all the transformations are clearly one-to-one, the uniqueness of this solution should follow from the uniqueness of the solution to equation (\ref{eq:df3}). It is easy to see that the solution to this equation is unique, as the difference of two possible solutions, with the same initial condition, has zero derivative, which implies that the two solutions are in fact equal.
\end{proof}
\noindent
The result of Proposition~\ref{prop:betweenJumps} tells us that one can calculate $g(s,t)$, for $T\leq t < T_n$, where $T_n$ is the first jump time of $N_t$ after $T$. In order to completely solve the filtering problem, one further has to solve the equation (\ref{eq:dgstCIR}) at jump times. This is the topic of the next section, where a recursive solution will be obtained for the case in which $\lambda_0$ has a Gamma distribution.

\section{Filtering at Jump Times and a General Solution}
\label{sec:atJumps}
In the previous section the filtering problem has been solved between jumps, for an arbitrary initial condition $w(s)$ for $g(s,t)$, at time $T>0$. In this section the filtering problem is solved at jump times, first for Example~\ref{ex:toy}, and after that for the case where the intensity follows the CIR model.
\begin{example}[Example~\ref{ex:toy} (continued)]
\label{ex:toy2}
\emph{
At jumps one obtains from equation (\ref{eq:dgstCIR})
\[\Delta g(s,t) = \left(\frac{\dd{s}{1}g(s,t-)}{\hat \lambda_{t-}^0}- g(s,t-)\right)\Delta N_t.\]
From this identity it easily follows that at a jump time $T>0$:
\begin{equation}
\label{eq:solutionAtJumps}
g(s,T) = \left(\hat\lambda_{T-}^0\right)^{-1} \dd{s}{1}g(s,T-).
\end{equation}
Combining the results between jumps and at jumps, one can obtain the solution to equation
\[\dif g(s,t) = -\dd{s}{1}g(s,t)\dif t + \left(\frac{\dd{s}{1}g(s,t-)}{\hat \lambda_{t-}^0}- g(s,t-)\right)\dif N_t.\]
At each jump time $T_n$, one has to take the derivative of the function $g(s,t)$, and divide by $\lambda_{T_n-}^0$; the resulting function can then be used as initial condition for the interval $[T_n,T_{n+1})$. Using an initial condition $g(s,0) = w(s)$, one obtains the solution
\[g(s,t) = w^{(N_t)}(s-t)\prod_{n=1}^{N_t} \left(\hat \lambda_{T_n-}^0\right)^{-1},\]
where $w^{(n)}(s)$ denotes the $n$-th derivative of $w(s)$. The conditional moment generating function is found from (\ref{eq:mgfFromGst}), and is given by
\[ \hat f(s,t) = \frac{g(s,t)}{g(0,t)}= \frac{w^{(N_t)}(s-t)}{w^{(N_t)}(-t)}.\]
If one assumes that $\lambda_0 \sim \Gamma(\alpha, \beta)$, one has
\begin{align}
f(s,0) &= \hat f(s,0) = \left(\frac{\beta}{\beta- s}\right)^{\alpha}, & \hat f(s,t) &= \left(\frac{\beta + t}{\beta+t - s}\right)^{\alpha + N_t}.
\end{align}
From this follows that at time $t>0$, $\lambda_t$ given $\F_t^N$ is distributed according to $\Gamma(\alpha + N_t, \beta +t)$. Further $\hat \lambda_t$ can easily be derived by a differentiation with respect to $s$:
\[\hat \lambda_t = \left.\dd{s}{1} \hat f(s,t)\right\vert_{s=0} = \frac{\alpha+N_t}{\beta + t}.\]}
\end{example}
\noindent
The solution in this example was easy to find, which could be expected, since $\lambda_t$ is constant over time in this case. The general Cox-Ingersoll-Ross model for the intensity is more complicated, but in the remainder of this section, also this problem is solved. At jumps one has the same equation as in Example~\ref{ex:toy2}, which is already solved in (\ref{eq:solutionAtJumps}). In Theorem~\ref{thm:CIRgeneralSol} the solution for $g(s,t)$ for the CIR model is given. Before this theorem is stated some notation is introduced.\\
Let $x,y \in \R$, $t\geq 0$ and put
\begin{align}
\label{equation:CIRAxty} A(x,t,y) &:= x\left((\tau - \alpha)e^{-\tau t} + \tau + \alpha\right) + 2 y \left( 1 - e^{-\tau t}\right)\\
\label{equation:CIRBst} B(s,t) &:= \rho s\left(e^{-\tau t }-1\right) + (\tau - \alpha)e^{-\tau t} + \tau + \alpha\\
\label{equation:CIRCxty}  C(x,t,y) &:= y \left((\alpha + \tau)e^{-\tau t} + \tau - \alpha\right) + \rho x \left(1 - e^{-\tau t}\right).
\end{align}
This notation allows us to write the general solution between jumps, (\ref{eq:solutionBetweenJumps}), as
\begin{equation}
\label{equation:CIRgeneralSolG}
g(s,t) = e^{\theta(\alpha - \tau)(t-T)} \left(\frac{2\tau}{B(s,t-T)}\right)^{2\theta}w \left(\frac{C\left(-\frac{2}{\rho},t-T,s\right)}{B(s,t-T)}\right)
\end{equation}
Next let $T_1,T_2,\ldots$ denote the jump times, and let $T_0 = 0 $. Then introduce the following notation:
\begin{align}
\label{equation:CIRAtT0} \A(t,T_0) &:= A(\phi, t, 1)&& \mbox{for } 0\leq t < T_1,\\
\label{equation:CIRAtTn} \A(t,T_n) &:= A(\A(T_n,T_{n-1}), t - T_n, \C(T_n,T_{n-1})) && \mbox{for } T_n \leq t < T_{n+1},\\
\label{equation:CIRCtT0} \C(t,T_0) &:= C(\phi, t, 1)&& \mbox{for } 0\leq t < T_1,\\
\label{equation:CIRCtTn} \C(t,T_n) &:= C(\A(T_n,T_{n-1}),t-T_n,\C(T_n,T_{n-1})) && \mbox{for } T_n \leq t < T_{n+1}.
\end{align}
With this notation, the main result of this paper can be stated. A recursive solution to the filtering problem is obtained, for the case where $\lambda_0$ has a Gamma distribution.
\begin{thm}
\label{thm:CIRgeneralSol}
Let $\lambda_0 \sim \Gamma(2\theta, \phi)$, for $\phi >0$ and $\theta = \frac{\mu_0\alpha}{\rho}>0$. Then one has
\[\hat f_0(s) = g(s,0) = \left(\frac{\phi}{\phi - s}\right)^{2\theta},\]
which is the moment generating function of the $\Gamma(2\theta,\phi)$ distribution. With the notation introduced in (\ref{equation:CIRAxty})-(\ref{equation:CIRCxty}) and (\ref{equation:CIRAtT0})-(\ref{equation:CIRCtTn}) one further has, for $T_n \leq t < T_{n+1}$,
\begin{equation}
\label{equation:CIRspecificSol}
g(s,t) = K(t) p_n(s,t) \left(\frac{1}{\A(t,T_n) - s \C(t,T_n)}\right)^{2\theta + n},
\end{equation}
where $p_0(s,t) \equiv 1$, and for $n\geq 1$, $p_n(s,t)$ is a polynomial of degree $n$ in s, that satisfies the recursion,
\begin{align}
\nonumber p_n(s,t) &= B^n(s,t-T_n)
\Bigg[p_{n-1}\left( \frac{C\left(-\frac{2}{\rho},t-T_n,s\right)}{B(s,t-T_n)},T_n\right)(2\theta + n - 1) \C(T_n,T_{n-1})\\
&\quad +
    \partial_1
    \left(
        p_{n-1}
        \left(
            \frac{C
            \left(
                -\frac{2}{\rho},t-T_n,s
            \right)}
            {B(s,t-T_n)}
            ,T_n
        \right)
    \right)
    \left(\A(T_n,T_{n-1}) - \frac{C\left(-\frac{2}{\rho},t-T_n,s\right)}{B(s,t-T_n)} \C(T_n,T_{n-1})
    \right), \Bigg]
\label{eq:pnst}
\end{align}
where $\partial_1$ denotes the derivative with respect to the first argument of $p_n$, and
\begin{equation}
\label{eq:kt}
K(t) = e^{\theta(\alpha - \tau)t} (2\tau\phi)^{2\theta}\prod_{m\geq 1,\ T_m \leq t} \left(\frac{(2\tau)^{2\theta}}{\hat \lambda_{T_m-}^0}\right).
\end{equation}
\end{thm}
\noindent In the proof of this theorem the following lemma is used.
\begin{lem}
\label{lemma:CIRsubCalcs}
With the notation from (\ref{equation:CIRAxty})-(\ref{equation:CIRCxty}) and (\ref{equation:CIRAtT0})-(\ref{equation:CIRCtTn}), the following relations hold for $n\geq 1$ and $x,y\in \mathbb{R}$:
\begin{itemize}
\item[(i)] $\A(T_n,T_n) = 2\tau \A(T_n,T_{n-1})$
\item[(ii)] $\C(T_n,T_n) = 2\tau \C(T_n,T_{n-1})$
\item[(iii)] $ x B(s,t) - yC\left(-\frac{2}{\rho},t,s\right) = A(x, t, y) - s C(x,t,y)$.
\end{itemize}
\end{lem}
\begin{proof}
\begin{itemize}
\item[(i)] From equations (\ref{equation:CIRAtTn}) and (\ref{equation:CIRAxty}) follows that
\begin{align*}
\A(T_n,T_n) &= A(\A(T_n,T_{n-1}),0,\C(T_n,T_{n-1}))\\
&= \A(T_n,T_{n-1})\left((\tau - \alpha)e^0 + \tau + \alpha\right) + \C(T_n,T_{n-1})\left(1 - e^0\right)\\
&= 2\tau \A(T_n,T_{n-1}).
\end{align*}
\item[(ii)] This follows along the same lines as in (i), using equations (\ref{equation:CIRCtTn}) and (\ref{equation:CIRCxty}).
\item[(iii)] Using equations (\ref{equation:CIRAxty}), (\ref{equation:CIRBst}) and (\ref{equation:CIRCxty}) one finds:
\begin{align*}
x B(s,t) - yC\left(-\frac{2}{\rho},t,s\right) &= x\left(\rho s\left(e^{-\tau t }-1\right) + (\tau - \alpha)e^{-\tau t} + \tau + \alpha\right) \\
&\quad- y \left(s \left((\alpha + \tau)e^{-\tau t} + \tau - \alpha\right) + 2 \left(1 - e^{-\tau t}\right)\right)\\
&= x\left((\tau - \alpha)e^{-\tau t} + \tau + \alpha\right) + 2y\left(1 - e^{-\tau t}\right)\\
&\quad- s\left(y\left((\alpha + \tau)e^{-\tau t} + \tau - \alpha\right) + x\rho \left(1- e^{-\tau t}\right)\right)\\
&= A(x,t,y) - s C(x,t,y).
\end{align*}
\end{itemize}
\end{proof}
\noindent Now, Theorem~\ref{thm:CIRgeneralSol} can be proved.
\begin{proof}[Proof of Theorem~\ref{thm:CIRgeneralSol}]
For each $n$ it has to be shown that (\ref{equation:CIRspecificSol}) holds at $T_n$, and between $T_n$ and $T_{n+1}$. First this is shown for $n=0$. Then the induction step is proved for $n\geq 1$.\\
$\mathbf{n=0}$: For $t = T_0 = 0$ one has by assumption:
\[g(s,0) = \left(\frac{\phi}{\phi - s}\right)^{2\theta}.\]
From (\ref{equation:CIRspecificSol}) one finds:
\begin{align*}
g(s,0) &= K(0) p_0(s,0) \left(\frac{1}{\A(0,0) - s \C(0,0)}\right)^{2\theta}\\
&= e^0 (2\tau \phi)^{2\theta} \left(\frac{1}{A(\phi,0,1) - s C(\phi,0,1)}\right)^{2\theta}\\
&= \left(\frac{2\tau \phi}{2\tau \phi - 2\tau s}\right)^{2\theta} = \left(\frac{\phi}{\phi - s}\right)^{2\theta}.
\end{align*}
Next the interval up to the first jump time, $0 < t < T_1$, is considered. From (\ref{equation:CIRgeneralSolG}) and the expression for $w(s) = g(s,0)$, one finds:
\begin{align*}
g(s,t) &= e^{\theta (\alpha - \tau)t} \left(\frac{2\tau}{B(s,t)}\right)^{2\theta}\left(\frac{\phi}{\phi - \frac{C\left(-\frac{2}{\rho},t,s\right)}{B(s,t)}}\right)^{2\theta}\\
&=e^{\theta (\alpha - \tau)t}  (2\tau \phi)^{2\theta} \left(\frac{1}{B(s,t)\phi - C\left(-\frac{2}{\rho},t,s\right)}\right)^{2\theta}\\
&= K(t) p_0(s,t) \left(\frac{1}{\A(t,0)- s \C(t,0)}\right)^{2\theta},
\end{align*}
which is the same expression as in (\ref{equation:CIRspecificSol}) for $n=0$. The final step in the derivation above follows from Lemma~\ref{lemma:CIRsubCalcs} (iii), with $x = \phi$ and $y = 1$, together with the definition of $K(t)$ in (\ref{eq:kt}).\\

\noindent $\mathbf{n\geq 1}$: Now it remains to prove the induction step. Therefore one can assume that equation (\ref{equation:CIRspecificSol}) holds for $n-1$. It then remains to show that the equation holds for $n$, at $T_n$ and between $T_n$ and $T_{n+1}$. First the jump is considered. Thus one has to calculate the derivative of $g(s,t)$ with respect to $s$, and take the left limit in $t=T_n$, further the derivative is divided by $\hat \lambda_{T_n-}^0$. By (\ref{eq:solutionAtJumps}) one has
\begin{align}
\nonumber g(s,T_n) &= \left(\hat \lambda_{T_n-}^0\right)^{-1}\dd{s}{1} g(s,T_n-)\\
\nonumber &= \left(\hat \lambda_{T_n-}^0\right)^{-1} \dd{s}{1}\left(K(T_n-)p_{n-1}(s,T_n-)\left(\frac{1}{\A(T_n,T_{n-1})- s \C(T_n,T_{n-1})}\right)^{2\theta +n - 1}\right).
\end{align}
Calculating the derivative with respect to $s$, leads to
\begin{align}
\nonumber g(s,t) &= \left(\hat \lambda_{T_n-}^0\right)^{-1} K(T_n-) \Bigg[ p_{n-1}(s,T_n)(2\theta + n - 1)\C(T_n,T_{n-1})\\
\label{eq:prf:deriv} &\quad+ \dd{s}{1} p_{n-1}(s,T_n)\left(\A(T_n,T_{n-1}) - s \C(T_n,T_{n-1})\right)\Bigg]\left(\frac{1}{\A(T_n,T_{n-1}) - s \C(T_n,T_{n-1})}\right)^{2\theta+n}.
\end{align}
From Lemma~\ref{lemma:CIRsubCalcs} (i) and (ii) follows that for the denominator in (\ref{eq:prf:deriv}) one has
\[\A(T_n,T_{n-1}) - s \C(T_n,T_{n-1}) = (2\tau)^{-1}\left(\A(T_n,T_{n}) - s \C(T_n,T_{n})\right).\]
Hence (\ref{eq:prf:deriv}) can be written as
\begin{align}
\nonumber g(s,T_n) &= \left(\hat \lambda_{T_n-}^0\right)^{-1} K(T_n-)(2\tau)^{2\theta} (2\tau)^n\Bigg[p_{n-1}(s,T_n)(2\theta + n - 1)\C(T_n,T_{n-1}) \\
\label{eq:prf:2tau} &\quad+ \dd{s}{1} p_{n-1}(s,T_n)\left(\A(T_n,T_{n-1}) - s \C(T_n,T_{n-1})\right)\Bigg]\left(\frac{1}{\A(T_n,T_n) - s \C(T_n,T_n)}\right)^{2\theta+n}.
\end{align}
From (\ref{eq:kt}) it is easy to see that $K(T_n) = K(T_n-)\left(\hat \lambda_{T_n-}^0\right)^{-1}(2\tau)^{2\theta}$, and further one has $ 2\tau = B(s,0)= B(s,T_n-T_n)$. From this follows that (\ref{eq:prf:2tau}) can be written as
\begin{align*}
g(s,T_n) &=  K(T_n)B^n(s,T_n-T_n)\Bigg[p_{n-1}(s,T_n)(2\theta + n - 1)\C(T_n,T_{n-1}) \\
&\quad+ \dd{s}{1} p_{n-1}(s,T_n)\left(\A(T_n,T_{n-1}) - s \C(T_n,T_{n-1})\right)\Bigg]\left(\frac{1}{\A(T_n,T_n) - s \C(T_n,T_n)}\right)^{2\theta+n}.
\end{align*}
This can be simplified further using the definition of $p_n(s,t)$ as given in (\ref{eq:pnst}), together with the identity $C\left(-\frac{2}{\rho},0,s\right)=\tau s$. This results in
\begin{align*}
g(s,T_n) &=  K(T_n)p_n(s,T_n)\left(\frac{1}{\A(T_n,T_n) - s \C(T_n,T_n)}\right)^{2\theta+n},
\end{align*}
which is the required result at $t=T_n$. Finally one has to check that (\ref{equation:CIRspecificSol}) holds for $T_n<t<T_{n+1}$. For this one can use the general solution (\ref{equation:CIRgeneralSolG}) with initial condition $w(s) = g(s,T_n)$. One finds
\begin{align*}
g(s,t) &= e^{\theta(\alpha-\tau)(t-T_n)}\left(\frac{2\tau}{B(s,t-T_n)}\right)^{2\theta} e^{\theta(\alpha-\tau)T_n} (2\tau \phi)^{2\theta} \prod_{m\geq 1,\ T_m \leq T_n}\left(\frac{(2\tau)^{2\theta}}{\hat\lambda_{T_m-}^0}\right)\\
&\quad\times p_n\left(\frac{C\left(-\frac{2}{\rho},t-T_n,s\right)}{B(s,t-T_n)},T_n\right) \left(\frac{1}{\A(T_n,T_n)- \frac{C\left(-\frac{2}{\rho}, t-T_n,s\right)}{B(s,t-T_n)}\C(T_n,T_n)}\right)^{2\theta+n}.
\end{align*}
Simplifying this expression yields:
\begin{align*}
g(s,t) &= e^{\theta(\alpha-\tau)t}(2\tau\phi)^{2\theta}\left(\prod_{m=1}^n \left(\frac{(2\tau)^{2\theta}}{\hat\lambda_{T_m-}^0}\right)\right)(2\tau)^{2\theta}B^n(s,t-T_n) p_n\left(\frac{C\left(-\frac{2}{\rho},t-T_n,s\right)}{B(s,t-T_n)},T_n\right)\\
&\quad\times \left(\frac{1}{2\tau B(s,t-T_n)\A(T_n,T_{n-1})- 2\tau C\left(-\frac{2}{\rho}, t-T_n,s\right)\C(T_n,T_{n-1})}\right)^{2\theta+n}.
\end{align*}
An application of Lemma~\ref{lemma:CIRsubCalcs}, with $x=\A(T_n,T_{n-1})$ and $y=\C(T_n,T_{n-1})$, and the definitions of $\A(t,T_n)$ and $\C(t,T_n)$ in (\ref{equation:CIRAtTn}) and (\ref{equation:CIRCtTn}), together with the definition of $K(t)$ results in
\begin{align*}
g(s,t) = K(t) \frac{1}{(2\tau)^n} B^n(s,t-T_n) p_n\left(\frac{C\left(-\frac{2}{\rho},t-T_n,s\right)}{B(s,t-T_n)},T_n\right) \left(\frac{1}{ \A(t,T_n)-s\C(t,T_n)}\right)^{2\theta+n}.
\end{align*}
Next, with the definition of $p_n(s,T_n)$ from (\ref{eq:pnst}), evaluated in $t=T_n$, together with $C(x,0,y) = 2\tau y$ and $B(s,0) = 2\tau$ one rewrites this to
\begin{align*}
g(s,t) &= K(t)\left(\frac{1}{ \A(t,T_n)-s\C(t,T_n)}\right)^{2\theta+n}\frac{1}{(2\tau)^n}B^n(s,t-T_n)\\
&\quad\times (2\tau)^n \Bigg[ p_{n-1}\left(\frac{C\left(-\frac{2}{\rho},t-T_n,s\right)}{B(s,t-T_n)},T_n\right) (2\theta + n - 1)\C(T_n,T_{n-1})\\
&\quad+ \partial_1\left( p_{n-1}\left(\frac{C \left( - \frac{2}{\rho}, t-T_n, s\right)}{B(s,t-T_n)}, T_n\right)\right)
\left(\A(T_n,T_{n-1}) - \frac{C\left(-\frac{2}{\rho},t-T_n,s\right)}{B(s,t-T_n)} \C(T_n,T_{n-1})\right)\Bigg]\\
&= K(t) p_n(s,t)\left(\frac{1}{ \A(t,T_n)-s\C(t,T_n)}\right)^{2\theta+n}.
\end{align*}
In the final step the definition of $p_n(s,t)$ is used, this time evaluated in $t$, which concludes the proof of (\ref{equation:CIRspecificSol}). From the definition of $B(s,t)$ and $C(x,t,y)$, with $y=s$, which are both linear in $s$, follows that $p_n(s,t)$ is a polynomial of degree $n$ in $s$.
\end{proof}
\noindent
This theorem provides a recursive solution to equation (\ref{eq:dgstCIR}), in case $\lambda_0$ is distributed according to a $\Gamma(2\theta,\phi)$ distribution. From (\ref{eq:mgfFromGst}) it already known that the conditional moment generating function can easily be obtained from an expression for $g(s,t)$. Now this has been found, the conditional moment generating function $\hat f(s,t)$ can be obtained easily.
\begin{cor}
\label{cor:CIRspecificSolF}
Under the assumptions of Theorem~\ref{thm:CIRgeneralSol} the conditional moment generating function $\hat f(s,t)$, for $T_n \leq t < T_{n+1}$, can be expressed as:
\begin{equation}
\label{eq:condMGF}
\hat f(s,t) = q_n(s,t)\left(\frac{Q(t,T_n)}{Q(t,T_n) - s}\right)^{2\theta + n},
\end{equation}
where
\begin{align*}
q_n(s,t) = \frac{p_n(s,t)}{p_n(0,t)}\textrm{ and } Q(t,T_n) = \frac{\A(t,T_n)}{\C(t,T_n)}.\hfill
\end{align*}
Here $q_n(s,t)$ is a polynomial of degree $n$ in $s$.
\end{cor}
\begin{proof}
The result follows directly from equation (\ref{eq:mgfFromGst}), Theorem~\ref{thm:CIRgeneralSol} and the definitions of $q_n$ and $Q$.
\end{proof}
\noindent
With the derivation of the conditional moment generating function the filtering problem has been solved, and one is able to calculate conditional default probabilities using the results in Example~\ref{ex:survival2}. To conclude this section it is observed that the conditional moment generating function in (\ref{eq:condMGF}) corresponds to a mixture of Gamma distributions.
\begin{remark}
\emph{
Corollary~\ref{cor:CIRspecificSolF} provides an expression for $\hat f(s,t)$ that involves the polynomial $q_n(\cdot,t)$. Deriving an explicit expression for $q_n(s,t)=p_n(s,t)/p_n(0,t)$ for any $n\geq0$ is quite complicated, but we can write
\[q_n(s,t) = \sum_{i=0}^n R_i^n(t)s^i,\]
where the coefficients, $R_i^n(t)$, of the polynomial follow directly from the coefficients of the polynomial in $s$, $p_n(s,t)$, which in turn can be obtained using the recursion (\ref{eq:pnst}).\\
Next, one can consider $n+1$ independent random variables $\Gamma_i$, where $\Gamma_i \sim \Gamma(2\theta + n - i, Q(t,T_n))$, for $i=0,1,\ldots,n$. Further, consider the discrete random variable $M^n$, independent of the $\Gamma_i$, which assumes the values $0,1,\ldots,n$, with probabilities $\pi_i^n(t)$, and define the random variable
\[X_t^n = \sum_{i=0}^n 1_{\{M=i\}} \Gamma_i.\]
The moment generating function of $X_t^n$ can easily be found, as $\Gamma_i$ and $M^n$ are independent, hence
\begin{align}
\label{eq:mgfX} \E\left[e^{sX_t^n}\right] = \sum_{i=0}^n\E\left[e^{s\Gamma_i}1_{\{M=i\}}\right]= \sum_{i=0}^n \pi_i^n(t) \E\left[e^{s\Gamma_i}\right]  = \sum_{i=0}^n \pi_i^n(t) \left(\frac{Q(t,T_n)}{Q(t,T_n) - s}\right)^{2\theta + n - i}.
\end{align}
The goal is to show that, by choosing the probabilities correctly, the moment generating function of $X_t^n$ equals the conditional moment generation function $\hat f(s,t)$. Therefore (\ref{eq:mgfX}) is first rewritten as
\begin{align}
\nonumber \E\left[e^{sX_t^n}\right] &=\left(\frac{Q(t,T_n)}{Q(t,T_n) - s}\right)^{2\theta + n}\sum_{i=0}^n \pi_i^n(t)\left(\frac{Q(t,T_n)-s}{Q(t,T_n)}\right)^i.
\end{align}
To have that both moment generating functions $\hat f(s,t)$ and (\ref{eq:mgfX}) are equal, it is required that
\[q_n(s,t) = \sum_{i=0}^n R_i^n(t)s^i = \sum_{i=0}^n \pi_i^n(t) \left(\frac{Q(t,T_n)-s}{Q(t,T_n)}\right)^i.\]
The right hand side of this equation can be written as
\begin{align*}
\sum_{i=0}^n \pi_i^n(t) Q(t,T_n)^{-i} \sum_{j=0}^i \left(\begin{array}{c}i\\j\end{array}\right)Q(t,T_n)^{i-j} s^j(-1)^j.
\end{align*}
This equation can be turned into a polynomial in $s$, by interchanging the summations, which leads to
\begin{align*}
\sum_{j=0}^n\sum_{i=j}^n \left(\begin{array}{c}i\\j\end{array}\right) \pi_i^n(t)Q(t,T_n)^{-j}s^j (-1)^j &=\sum_{j=0}^n s^j \left((-1)^j Q(t,T_n)^{-j}\sum_{i=j}^n \left( \begin{array}{c}i\\j\end{array} \right)\pi_i^n(t) \right).
\end{align*}
The moment generating functions are equal when
\[R_j^n(t) = (-1)^j Q(t,T_n)^{-j}\sum_{i=j}^n \left( \begin{array}{c}i\\j\end{array} \right)\pi_i^n(t),\]
for $j=0,1,\ldots,n$. This can be solved iteratively, starting from $j=n$, which results in the probabilities
\begin{align}
\label{eq:mixProbs}
\pi_j^n(t) &= (-1)^j R_j^n(t) Q(t,T_n)^j - \sum_{i=j+1}^n \pi_i^n(t)\left( \begin{array}{c}i\\j\end{array} \right).
\end{align}
It is not immediately clear from (\ref{eq:mixProbs}) that the $\pi_j^n(t)$ are all non-negative and sum to one. It turns out however that this is indeed the case for $T_n \leq t < T_{n+1}$, which means that the $\pi_j^n(t)$ can be interpreted as probabilities. It is however far from trivial to provide a general proof for all $n\geq 0$. We confine ourselves to illustrate this fact by some examples. In figure~\ref{fig:probs}, two graphs are given in which the probabilities are plotted.
}
\begin{figure}[hbt]
\subfigure[][{$\pi_0^2$ (short dashed line), $\pi_1^2$ (long dashed line) and $\pi_2^2$ (solid line)}]{
\includegraphics[width = 0.43\linewidth]{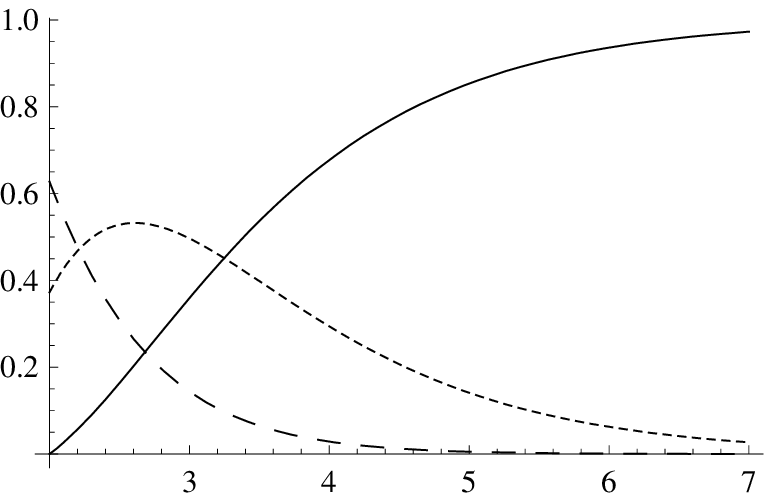}
\label{fig:probs2}}
\hspace{0.02\linewidth}
\subfigure[][{$\pi_0^3$ (dot dashed line), $\pi_1^3$ (short dashed line), $\pi_2^3$ (long dashed line) and $\pi_3^3$ (solid line)}]{
\includegraphics[width = 0.43\linewidth]{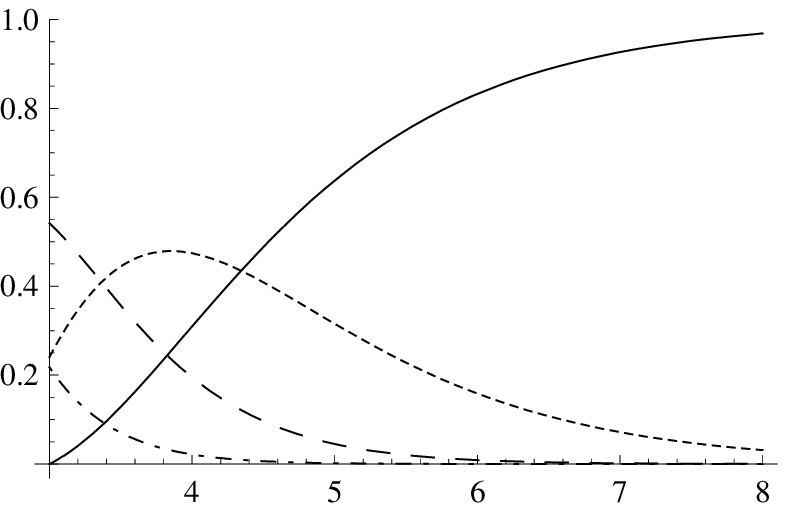}
\label{fig:probs3}}
\caption{Graphs of the mixing probabilities after two jumps of the process $N_t$, (a), and after three jumps, (b). The values of the previous jump times, $T_1$ and $T_2$ in case (a), and $T_1,\ T_2$ and $T_3$ in case (b), are taken as $T_i = i$, such that one is able to calculate the $\pi_j^n(t)$. The model parameters are chosen to be $\alpha = 0.5$, $\beta = 0.5$, $\mu_0 = 0.4$ and $\phi = 4.0$.}
\label{fig:probs}
\end{figure}
\end{remark}
\newpage
\bibliographystyle{alpha}
\bibliography{references,thesisRefs}

\end{document}